\documentclass[12pt]{article}
\usepackage{amsfonts}
\usepackage{graphicx}
\usepackage{amsmath}
\usepackage{float}
\newcommand {\be}{\begin{equation}}
 \newcommand {\ee}{\end{equation}}
 \newcommand {\bea}{\begin{array}}
 
 \newcommand {\eea}{\end{array}}

\textwidth=6.5in \hoffset=-.75in \voffset=-1in \numberwithin{equation}{section}
\numberwithin{figure}{section}

\begin{document}

\begin{titlepage}
\vspace{1cm} 
\begin{center}
{\Large \bf {Hidden and Generalized Conformal Symmetry of Kerr-Sen Spacetimes}}\\
\end{center}
\vspace{2cm}
\begin{center}
A. M. Ghezelbash{\footnote{masoud.ghezelbash@usask.ca}}, H. M. Siahaan{\footnote{hms923@mail.usask.ca}}
\\
Department of Physics and Engineering Physics, \\ University of Saskatchewan, \\
Saskatoon, Saskatchewan, S7N 5E2, Canada\\
\vspace{1cm}
\end{center}

\begin{abstract}
It is recently conjectured that generic non-extremal Kerr black hole could be holographically dual to a hidden conformal field theory in two dimensions. Moreover, it is known that there are two CFT duals (pictures) to describe the charged rotating black holes which correspond to angular momentum $J$ and electric charge $Q$ of the black hole. Furthermore these two pictures can be incorporated by the CFT duals (general picture) that are generated by $SL(2,\mathbb{Z})$ modular group.
The general conformal structure can be revealed by looking at charged scalar wave equation in some appropriate values of frequency and charge. 
In this regard, we consider the wave equation of a charged massless scalar field in background of Kerr-Sen black hole and show in the ``near region", the wave equation can be reproduced by the Casimir operator of a local $SL(2,\mathbb{R})_L \times SL(2,\mathbb{R})_R$ hidden conformal symmetry. We can find the exact agreement between macroscopic and microscopic physical quantities like entropy and absorption cross section of scalars for Kerr-Sen black hole. 
We then find an extension of vector fields that in turn yields an extended local family of $SL(2,\mathbb{R})_L \times SL(2,\mathbb{R})_R$ hidden conformal symmetries, parameterized by one parameter. For some special values of the parameter, we find a copy of $SL(2,\mathbb{R})$ hidden conformal algebra for the charged  Gibbons-Maeda-Garfinkle-Horowitz-Strominger black hole in the strong deflection limit.
\end{abstract}
\end{titlepage}\onecolumn 
\bigskip 

\section{Introduction}
\label{sec:intro}
According to conjectured Kerr/CFT correspondence, the physical properties of an extremal Kerr black hole could be related to properties of a conformal field theory. More explicitly, the microscopic entropy and near-super radiant modes of four-dimensional extremal Kerr black hole can be derived by studying the dual chiral conformal field theory associated with the diffeomorphisms of near horizon geometry of the Kerr black hole \cite{stro}.   
  
The Kerr/CFT correspondence has been studied extensively for different four and higher dimensional extremal rotating black holes which the dual chiral conformal field theory always contains a left-moving sector \cite{KerrCFT1,KerrCFT2,KerrCFT3}. 
For these extremal black holes, the near horizon geometry contains a copy of AdS space with isometries that could be extended to Virasoro algebra, hence it may explain the appearance of conformal structure. 

However, the standard techniques of Kerr/CFT correspondence for extremal rotating black holes can not be applied to non-extremal black holes because there is no simple symmetry near the non-extremal black hole horizon that may point to conformal structure. Moreover for non-extremal black holes, the right-moving sector of dual CFT turns on and there is no consistent boundary conditions that allow for both left and right-moving sectors in CFT.

However, as it is noted in \cite{cms}, there is other conformal invariance, known as hidden conformal invariance, in the solution space of the wave equation in background of rotating non-extremal black holes. This means the existence of conformal invariance in a near horizon geometry is not a necessary condition, and the hidden conformal invariance is sufficient to have a dual CFT description. The idea of hidden conformal symmetry in the solution space of wave equation for a neutral scalar field in different rotating backgrounds was explored in detail in \cite{othe,KSChen}. 

For the class of four-dimensional rotating charged black holes such as Kerr-Newman, there are two dual CFTs; one associated with the rotation of black hole while the other one is associated with the electric charge of black hole. The two different dual pictures of black hole are called J and Q pictures, respectively \cite{Chen-JQpic}.
The angular momentum and the charge of Kerr-Newman black hole are in correspondence with the rotational symmetry of black hole in $\phi$ direction and the gauge symmetry, respectively. The latter symmetry can be considered geometrically as the rotational symmetry of the uplifted Kerr-Newman black hole in the fifth-direction $\chi$. As a result, the combination of two rotational symmetries of uplifted five-dimensional Kerr-Newman black hole lead to two new CFTs ($\phi '$ and $\chi '$ pictures \cite{chen-SL2Z}). Moreover, these two pictures neatly can be embedded into a general picture by using the torus $(\phi,\chi)$ modular group $SL(2,\mathbb{Z})$.

One can easily obtain the CFT results in J and Q pictures for Kerr-Newman black hole \cite{Chen-JQpic} as special case of the CFT results in general picture where the transformation is given by the unit element of group $SL(2,\mathbb{Z})$.

One other class of rotating charged black hole solutions in four dimensions is Kerr-Sen geometry \cite{sen}. The solution includes three non gravitational fields: an antisymmetric tensor field, a vector field and a dilaton. The Kerr-Sen solution is an exact solutions to the equations of motion of effective action of heterotic string theory in four dimensions. In \cite{GH2}, it was shown that for extremal Kerr-Sen black hole, the central charges of dual chiral CFT don't get any contributions from the non-gravitational fields. 
Moreover, the central charges lead to the microscopic entropy of black hole that is in perfect agreement with Bekenstein-Hawking entropy. We also notice the Kerr-Sen solutions contain a scalar dilatonic field, but the solution space of dilaton equation does not show any conformal symmetry. This in turn is in the same direction and in agreement with previously observation of no contribution of non-gravitational fields to the central charges of dual CFT in extremal case \cite{GH2}.

Inspired by existence of different CFT pictures for the four-dimensional non-extremal rotating charged Kerr-Newman black hole, 
the main motivation for this article is to investigate the CFT results in general picture and so the possibility of finding the CFT results in $\phi '$ and $\chi '$ pictures 
for the generic non-extremal Kerr-Sen black hole.

In this regard, we consider the equation of motion for a charged scalar field in the background of Kerr-Sen black hole and look for the hidden conformal symmetry in general picture. The charge of scalar field appears to be crucial in determining the existence of general picture, hence we can not consider the wave equation of a neutral scalar field as in \cite{KSChen}. 
We then discuss the absorption of scalar fields in the near region of non-extremal Kerr-Sen black hole. 

Moreover we find an extended version of hidden conformal generators \cite{Lowe} that involve one parameter for the class of Kerr-Sen solutions. These conformal generators in the appropriate limits, provide a completely new set of conformal symmetry generators for the charged Gibbons-Maeda-Garfinkle-Horowitz-Strominger black hole.  

The article is organized as follows: In section \ref{sec:sec2scalarfield}, we consider the  wave equation of a massless charged scalar field in the background of Kerr-Sen spacetimes. 
We show in some appropriate limits of parameters and using the general $SL(2,\mathbb{Z})$ modular transformation, the equation of motion can be simplified in the near region of Kerr-Sen black hole. 

In section \ref{sec:sec3hidden}, we show the radial part of wave equation in the near region, can be written in terms of $SL(2,\mathbb{R})_L \times SL(2,\mathbb{R})_R$ Casimir operators in $\phi '$ picture. Moreover, we find the microscopic entropy of the dual CFT and compare it to the macroscopic Bekenstein-Hawking entropy of the Kerr-Sen spacetimes. 

In section \ref{sec:sec4:absorption}, we calculate the absorption cross section of scalars in the near region of Kerr-Sen black hole and show explicitly the result is in perfect agreement with the finite temperature absorption cross section for a two-dimensional conformal field theory. 

In section \ref{sec:sec5kappa}, we introduce the deformed equation of motion for the test field and find explicitly two classes of generators that generate a generalized hidden conformal symmetry for the Kerr-Sen black hole. The generators can be used to find the hidden conformal symmetry for the charged Gibbons-Maeda-Garfinkle-Horowitz-Strominger black hole. In section \ref{sec:concl}, we wrap up the article with conclusions, few comments and open questions.  In general, our results in this paper provide further supporting evidence for the (generalized) hidden conformal symmetry of Kerr-Sen spacetimes as well as a new set of conformal symmetry generators for the charged Gibbons-Maeda-Garfinkle-Horowitz-Strominger black hole.  
 
\section{The Charged Scalar Field in Background of Kerr-Sen Spacetimes }\label{sec:sec2scalarfield}
The Kerr-Sen solution \cite{sen} is an exact classical four dimensional black hole solution in the low energy heterotic string field theory. In the Boyer-Lindquist coordinates, the Kerr-Sen metric can be rewritten as 
\begin{eqnarray}
ds^2&=&-(1-\frac{2Mr}{\rho^2})dt^2+\rho^2(\frac{dr^2}{\Delta}+d\theta^2)\nonumber\\
&-&\frac{4Mra}{\rho^2}\sin^2\theta dtd\phi+\{r(r+2b)+a^2+\frac{2Mr a^2\sin^2\theta}{\rho^2}\}\sin^2\theta d\phi^2,\label{intro1}
\end{eqnarray}
where $\rho^2=r(r+2b)+a^2\cos^2\theta$, $\Delta=r(r+2b)-2Mr+a^2$, and $b=Q^2/2M$. The non-gravitational fields: the dilaton, gauge field and the antisymmetric tensor field are given respectively by \cite{KS1,KS2}
\begin{eqnarray}
D&=&-\frac{1}{2}\ln \frac{{r}({r}+2b)+a^2\cos^2\theta}{{r}^2+a^2\cos^2\theta},\label{dil}\\
A_{{t}}&=&\frac{-{r}Q}{\rho ^2},\label{A1}\\
A_{{\phi}}&=&\frac{{r}Qa\sin^2\theta}{\rho ^2},\label{A2}\\
B_{{t}{\phi}}&=&\frac{b{r}a\sin^2\theta}{\rho ^2}.\label{Anti}
\end{eqnarray}
The outer horizon of black hole is located at $r_+=M-b+\sqrt{(M-b)^2-a^2}$, while the Hawking temperature, angular velocity of horizon and electrostatic potential are given by
\begin{eqnarray}
T_H&=&\frac{\sqrt{(2M^2-Q^2)^2-4J^2}}{4\pi M(2M^2-Q^2+\sqrt{(2M^2-Q^2)^2-4J^2})},\label{TH} \\
\Omega_H&=&\frac{J}{M(2M^2-Q^2+\sqrt{(2M^2-Q^2)^2-4J^2})},\label{Omeg}\\
V _H  &=& Q/2M.\label{electricpot}
\end{eqnarray}
For $b=0$, all non-gravitational fields (\ref{dil})-(\ref{Anti}) vanish and metric (\ref{intro1}) changes simply to the metric of Kerr black hole. For  generic non-zero $b$, the Kerr-Sen solution (\ref{intro1}) (along with the non-gravitational fields (\ref{dil})-(\ref{Anti})) is an interesting gravitational system in the context of string theory; quite different from Kerr solution in general relativity. The Kerr-Sen black hole (\ref{intro1}) approaches to the metric of charged Gibbons-Maeda-Garfinkle-Horowitz-Strominger black hole in the strong deflection limit \cite{bharda} where the rotational parameter $a\rightarrow 0$. 

We consider a massless scalar field $\Phi$ with charge $e$ as a probe in background (\ref{intro1}). The minimally coupled Klein-Gordon equation for the massless scalar field $\Phi$ is  
\begin{equation}
    \left( {\nabla _\mu   - ieA_\mu  } \right)\left( {\nabla ^\mu   - ieA^\mu  } \right)\Phi  = 0,\label{eq1}
\end{equation}
where $A_\mu$ is given by (\ref{A1}) and (\ref{A2}). 
We separate the coordinates in scalar wave function as
\begin{equation}
\Phi \left( {r,t,\theta ,\phi } \right) = \exp \left( {im\phi  - i\omega t} \right)S\left( \theta  \right)R\left( r \right),\label{eq2}
\end{equation}
where the radial and angular functions $R(r)$ and $S(\theta)$ are solutions to radial equation
\begin{equation}
    \partial _r \left( {\Delta \partial _r R\left( r \right)} \right) + \left( {\frac{{\left( {\gamma r - ma} \right)^2 }}{\Delta } + \omega^2 \Delta  + 2\delta r - \sigma } \right)R\left( r \right) = 0,\label{scalar-rad}
\end{equation}
and angular equation
\begin{align}
	\frac{1}{{\sin \theta }}\partial _\theta  \left( {\left( {\sin \theta } \right)\partial _\theta  S\left( \theta  \right)} \right) + \left( {\sigma  - \frac{{m^2 }}{{\sin ^2 \theta }} - \omega ^2 a^2 \sin ^2 \theta } \right)S\left( \theta  \right) = 0,\label{eq4}
\end{align}
respectively. In  equation (\ref{scalar-rad}), $\gamma  = 2M\omega  - eQ$, $\delta  = \gamma \omega$ and $\sigma$ is the separation constant. We notice the radial equation (\ref{scalar-rad}) can be rearranged to 
\begin{align}
\partial _r \left( {\Delta \partial _r R\left( r \right)} \right) + \left( {\frac{{\left( {2M\omega r_ +   - eQr_ +   - ma} \right)^2 }}{{\left( {r - r_ +    } \right)\left( {r_ +   - r_ -  } \right)}} - \frac{{\left( {2M\omega r_ -   - eQr_ -   - ma} \right)^2 }}{{\left( {r - r_ -  } \right)\left( {r_ +   - r_ -  } \right)}} + f\left(r\right)} \right)R\left( r \right) = \sigma R\left( r \right),\label{eq5}
\end{align}
where $f\left( r \right) = \left( {\Delta  + 4M\left( {M + r} \right)} \right)\omega ^2  - \left( {2M + r} \right)2eQ\omega  + e^2 Q^2 $, and $r_ -   = M - b - \sqrt {\left( {M - b} \right)^2  - a^2 }$ is the inner horizon of black hole.
To simplify the equation of motion (\ref{eq5}), we consider the
low frequency scalar field $\omega  <  < 1/M $ and so in the near region geometry defined by $r <  < 1/\omega $ and with assumption that electric charge of scalar field satisfies $eQ<<1$\footnote{{From the definition of $\gamma$, we can acknowledge that $eQ$ has the same dimension as $M\omega$}.}, 
we can neglect $f(r)$ in the left hand side of (\ref{eq5}).

As we notice, the electric charge of scalar field $e$ couples to the black hole charge $Q$ in
the radial equation (\ref{eq5}). The existence of $eQ$ term in radial equation is necessary to investigate the dual CFTs in general picture. 
The lack of $eQ$ term in the radial equation of neutral scalar field (as in reference \cite{KSChen}) hinders the general picture of Kerr-Sen geometry.    

In fact, the Kerr/CFT correspondence calculation in general picture shows the electric charge of Kerr-Newman black hole as well as the angular momentum of black hole 
enters in the CFT quantities such as the central charges and the hidden conformal symmetry generators \cite{Chen-JQpic}. 
Indeed 
to realize the (hidden) conformal symmetry of charged rotating black holes in general picture, one must consider a charged scalar field.
An immediate result of CFT calculations in general picture is that
by looking at the dual CFT quantities, one can observe the presence of electric charge (hair) of the black hole. 
The authors in \cite{Chen-JQpic} proposed that each macroscopic hair of black holes, may be associates to a dual CFT. 
In general picture, we consider the $SL(2,\mathbb{Z})$ transformation for the torus generated by $\phi$ and $\chi$ coordinates, given by \cite{chen-SL2Z} transformation
\begin{equation}
\left( {\begin{array}{*{20}c}
   {\phi '}  \\
   {\chi '}  \\
\end{array}} \right) = \left( {\begin{array}{*{20}c}
   \alpha  & \beta   \\
   \eta  & \tau   \\
\end{array}} \right)\left( {\begin{array}{*{20}c}
   \phi   \\
   \chi   \\
\end{array}} \right),\label{SL2trans}
\end{equation} 
where the two $U(1)$ symmetries of black hole are associated with $\phi$ and $\chi$ coordinates. The first symmetry is simply the rotational symmetry of the black hole along $\phi$ direction. The second symmetry is associated to the rotational symmetry of the uplifted black hole into five-dimensions (the fifth coordinate is $\chi$) and in fact this symmetry is equivalent to the original gauge symmetry of the four-dimensional charged rotating black hole. 
Such a transformation doesn't change the phase of the charged scalar field (\ref{eq1}) with electric charge $e$;
$e^{im\phi  + ie\chi }  = e^{im'\phi ' + ie'\chi '} $ which yields 
$m = \alpha m' + \eta e',e = \beta m' + \tau e'$.
Consequently, {in $\phi '$ picture, 
the radial equation (\ref{eq5}) for low frequency massless charged scalar field in the near region of Kerr-Sen spacetime can be rewritten as
\begin{eqnarray}
\partial _r \left( {\Delta \partial _r R\left( r \right)} \right) &+& \left( {\frac{{\left( {2Mr_ +  \omega  - \left( {Qr_ +  \beta + a\alpha} \right)m' } \right)^2 }}{{\left( {r - r_ +  } \right)\left( {r_ +   - r_ -  } \right)}}} 
- {\frac{{\left( {2Mr_ -  \omega  - \left( {Qr_ -  \beta + a\alpha} \right)m' } \right)^2 }}{{\left( {r - r_ -  } \right)\left( {r_ +   - r_ -  } \right)}}} \right)R\left( r \right) \nonumber \\
&=& l\left( {l + 1} \right)R\left( r \right),
 \label{eq7}
\end{eqnarray}
where we have chosen the separation constant $\sigma  = l\left( {l + 1} \right)$.
Moreover, to get the $\chi '$ picture, we should turn off the momentum along $\phi '$ coordinate. In this case, the radial equation (\ref{eq5}) for low frequency massless charged scalar field in the near region of Kerr-Sen becomes the same as equation (\ref{eq7}) by replacing $\alpha$, $\beta$ and $m'$ to $\eta$, $\tau$ and $e'$ respectively.  
 
\section{Hidden Conformal Symmetry of Kerr-Sen Geometry in General Picture}\label{sec:sec3hidden}
In this section, we find the hidden conformal symmetry of the radial equation (\ref{eq7}) for the massless charged scalar field in the near region of Kerr-Sen black hole in general picture. We define $\omega^+, \omega^-$ and $y$ \textit{as the conformal coordinates} in terms of coordinates $t,r$ and $\phi '$ by
\begin{eqnarray}
 \omega^{+}&=&\sqrt{\frac{r-r_{+}}{r-r_{-}}}\exp(2\pi T_{R}\phi '+2n_{R}t),
  \label{intro14}\\
  \omega^{-}&=&\sqrt{\frac{r-r_{+}}{r-r_{-}}}\exp(2\pi T_{L}\phi '+2n_{L}t),
  \label{intro15}\\
  y&=&\sqrt{\frac{r_{+}-r_{-}}{r-r_{-}}}\exp(\pi( T_{R}+T_{L})\phi '+(n_{R}+n_{L})t).
\label{intro16}
 \end{eqnarray} 
In terms of conformal coordinates, we also define the right and left moving vector fields  
\begin{equation}
H_1=i\partial_{+},~~~~H_0=i(\omega^{+}\partial_{+}+\frac{1}{2}y
\partial_{y}),~~~~~H_{-1}=i((\omega^{+})^2\partial_{+}+\omega^{+}y\partial_{y}-y^2\partial_{-}),~~~~
\label{intro19}
\end{equation}
and
\begin{equation}
\overline{H}_1=i\partial_{-},~~~~\overline{H}_0=i(\omega^{-}\partial_{-}+\frac{1}{2}y\partial_{y}),
~~~~~\overline{H}_{-1}=i((\omega^{-})^2\partial_{-}+\omega^{-}y\partial_{y}-y^2\partial_{+}),
\label{intro20}
\end{equation}
respectively. 
The vector fields (\ref{intro19}) satisfy the $SL(2,\mathbb{R})$ algebra
\begin{equation}
 ~~[H_0,H_{\pm1}]=\mp i H_{\pm 1},~~~~~~~~[H_{-1},H_1]=-2iH_0,~~
\label{intro21}
\end{equation}
and similarly for $\overline{H}_1,\overline{H}_0$ and $\overline{H}_{-1}$. 
The quadratic Casimir operators of $SL(2,\mathbb{R})_R$ and $SL(2,\mathbb{R})_L$ with generators $H_{\pm 1},H_0$ and $\bar H_{\pm 1},\bar H_0$ respectively, are equal and in conformal coordinates are given by
 \begin{eqnarray}
H^2&=&-H_{0}^2+\frac{1}{2}(H_1H_{-1}+H_{-1}H_{1})\\
&=&\bar H^2=-\bar H_{0}^2+\frac{1}{2}(\bar H_1 \bar H_{-1}+\bar H_{-1} \bar H_{1})\\
&=&\frac{1}{4}
(y^2\partial_{y}^2-y\partial_{y})+y^2\partial_{+}\partial_{-}.
\label{intro28}
\end{eqnarray}
It is straightforward to show that in terms of coordinates $t,r,\theta,\phi '$ (see appendix), the Casimir operators (\ref{intro28}) 
reduce simply to the radial equation (\ref{eq7}) in $\phi '$ picture,
 \be \begin{array}{cc}
H^2R(r)=\bar{H}^2R(r)=l(l+1)R(r),\end{array} \label{intro29}\ee
by choosing the right and left temperatures $T_R$ and $T_L$ 
\begin{equation}
T_R  = \frac{{r_ +   - r_ -  }}{{4\pi a\alpha}},
T_L  = \frac{{r_ +   + r_ -  }}{{4\pi a\alpha}},
\label{intro17}
\end{equation}
and  \begin{equation} n_R  = - \frac{{\left( {r_ +   - r_ -  } \right)\beta Q}}{{8\alpha aM}}, n_L  = - \frac{{\left( {2a\alpha  + \left( {r_ +   + r_ -  } \right)\beta Q} \right)}}{{8\alpha aM}},\label{intro18}\end{equation}
where $\alpha$ and $\beta$ are the parameters of $SL(2,\mathbb{Z})$ modular transformation (\ref{SL2trans}).

As we notice, the temperatures of CFT dual to Kerr-Sen black hole in $\phi '$ picture depend only on $\alpha$, while $n_L$ and $n_R$ depend on both $\alpha$ and $\beta$. The dependence of CFT temperatures on $SL(2,\mathbb{Z})$ parameters for Kerr-Sen is different than Kerr-Newman black hole. In the latter case, the CFT temperatures in $\phi '$ picture depend on both parameters $\alpha$ and $\beta$.  The CFT temperatures (\ref{intro17}), $n_L$ and $n_R$ (\ref{intro18}) reduce to the results in J picture when $\alpha=1$ and $\beta=0$ \cite{KSChen}.   

In $\chi '$ picture, the radial equation is given by equation (\ref{eq7}) where one replaces $\alpha$, $\beta$ and $m'$ to $\eta$, $\tau$ and $e'$ respectively.
After changing to the conformal coordinates (\ref{intro14})-(\ref{intro16}) (with replacing $\phi '$ to $\chi '$), we find the quadratic Casimir operators of $SL(2,\mathbb{R})_R$ and $SL(2,\mathbb{R})_L$ reduce to the radial equation in $\chi '$ picture by choosing the right and left temperatures $T_R$ and $T_L$ as
\begin{equation}
T_R  = \frac{{r_ +   - r_ -  }}{{4\pi a\eta}},
T_L  = \frac{{r_ +   + r_ -  }}{{4\pi a\eta}},
\label{intro17chi}
\end{equation}
and  \begin{equation} n_R  = - \frac{{\left( {r_ +   - r_ -  } \right)\tau Q}}{{8\eta aM}}, n_L  = - \frac{{\left( {2a\eta  + \left( {r_ +   + r_ -  } \right)\tau Q} \right)}}{{8\eta aM}},\label{intro18chi}\end{equation}
where $\eta$ and $\tau$ are the parameters of $SL(2,\mathbb{Z})$ modular transformation (\ref{SL2trans}). 
We notice for unit element of $SL(2,\mathbb{Z})$ where $\eta=0$ and $\tau=1$, the temperatures are not finite that indicates the Q picture for the Kerr-Sen geometry is not well defined. The same type of calculation for Kerr-Newman black hole in $\chi '$ picture shows taking $\eta=0$ and $\tau=1$ leads to the well defined Q picture for the Kerr-Newman black hole \cite{Chen-JQpic}. The non-existent Q picture for the Kerr-Sen geometry  hinders uplifting the black hole into five dimensional spacetime, in contrast to Kerr-Newman black hole.  

We note that equation (\ref{intro29}) signals the existence of $SL(2,\mathbb{R})_L \times SL(2,\mathbb{R})_R$ hidden conformal symmetry in $\phi '$ picture for the Kerr-Sen black hole.
We should emphasise that $SL(2,\mathbb{R})_L \times SL(2,\mathbb{R})_R$ is only a local hidden conformal symmetry for the solution space of massless charged scalar field in near region of Kerr-Sen geometry. The local symmetry is generated by the vector fields (\ref{intro19}),(\ref{intro20}). The reason is these vectors in $\phi '$ picture are not periodic under $\phi ' \sim \phi ' +2\alpha\pi$ identification,  so they can't be defined globally. We may conclude the existence of local $SL(2,\mathbb{R})_L \times SL(2,\mathbb{R})_R$ hidden conformal symmetry in $\phi '$ picture, suggests that we assume the dynamics of the near region can be described by a dual CFT. To verify this assumption, we try to find the microscopic entropy of the dual CFT which according to the Cardy formula, is given by
 \be \begin{array}{cc}
S_{CFT}=\frac{\pi^2}3({c_{L}T_{L}+c_{R}T_{R}}),
\end{array} \label{intro30}\ee
where $T_R$ and $T_L$ are the CFT temperatures in $\phi$ picture, given by (\ref{intro17}). The central charges of dual CFT for extremal Kerr-Sen black holes were obtained in \cite{GH2} based on analysis of the asymptotic symmetry group. For the case of non-extremal black hole, we assume that the conformal symmetry connects smoothly to that of the extremal case; so we consider the central charges given by
 \be \begin{array}{cc}
c_{R}=c_{L}=12\alpha J.
\end{array} \label{intro31}\ee
We notice in the case of $\alpha=1$, (\ref{intro31}) reduces to $12J$ which is the central charge in the J-picture. 
The central charges (\ref{intro31}) and temperatures (\ref{intro17}) yield the microscopic entropy of CFT (\ref{intro30}) in $\phi '$ picture as
 \be \begin{array}{cc}
S_{CFT}=2\pi M r_{+},
\end{array} \label{intro32}
\ee
which is in complete agreement with the macroscopic Bekenstein-Hawking entropy of Kerr-Sen spacetime. The macroscopic Bekenstein-Hawking entropy of Kerr-Sen black hole is given by \cite{GH2,ALks}
\begin{equation}
S=\pi \left( {2M^2  - Q^2  + \sqrt {\left( {2M^2  - Q^2 } \right)^2  - 4J^2 } } \right), \label{entKS}
\end{equation}
which is equal to $S_{CFT}$ upon substitution $r_ +   = M - b + \sqrt {\left( {M - b} \right)^2  - a^2 }$, $J = aM$, and $b = Q^2 /2M$. 

\section{Absorption Cross Section of Near Region Scalars in $\phi '$ Picture}
\label{sec:sec4:absorption}
In this section, to further show that the dynamics of the near region can be described by a dual CFT in $\phi '$ picture, we consider the absorption cross section of scalars in the near region of Kerr-Sen spacetime. We find that the absorption cross section could be reproduced correctly by dual CFT. In this regard, we introduce the new coordinate $p$, given by \cite{malda-strom}
\begin{align}
p = \frac{{r - r_ +  }}{{r - r_ -  }}\label{p}.
\end{align}
By using the following relation that is obtained from (\ref{p}),
\begin{align}
\Delta \partial _r  = \left( {r_ +   - r_ -  } \right)p\partial _p ,\label{pp}
\end{align}
one can rewrite the radial part of Klein-Gordon equation (\ref{scalar-rad}) in terms of new coordinate $p$ as 
\begin{align}
p\left( {1 - p} \right)\partial _p^2 R\left( p \right) + \left( {1 - p} \right)\partial _p R\left( p \right) + \left( {\frac{{C_1 ^2 }}{p} - C_2 ^2  - \frac{{C_3 }}{{1 - p}}} \right)R\left( p \right) = 0,\label{eq-p}
\end{align}
where the constants $C_1$, $C_2$ and $C_3$ are  
\begin{eqnarray}
	C_1  &=& \left( {\frac{{2Mr_ +  \omega  - \left( {Qr_ +  \beta + a\alpha } \right)m'  }}{{r_ +   - r_ -  }}} \right),\label{c1}
\\
	C_2  &=& \left( {\frac{{2Mr_ -  \omega  - \left( {Qr_ -  \beta + a\alpha } \right)m'  }}{{r_ +   - r_ -  }}} \right),\label{c2}
\\
C_3 &=& l\left( {l + 1} \right).\label{c3}
\end{eqnarray}
The in-going solution for the equation (\ref{eq-p}) is
\begin{align}
	R_{in} \left( r \right) = C {\rm{ }}p^{ - iC_1 } \left( {p - 1} \right)^{ - l} {}_2F_1 \left( { - l - i\left( {C_1  - C_2 } \right), - l - i\left( {C_1  + C_2 } \right);1 - 2iC_1 ;p} \right),\label{Rin}
\end{align}
where $C$ is a constant of integration and $ {}_2 F_1 $ is the hypergeometric function. The in-going solution (\ref{Rin}) on the outer boundary of the matching region where $r>>M$ behaves as, 
\begin{align}
R_{in}  \sim Ar^l,\label{Rin-far}
\end{align}
where $ 
A = {}_2F_1 \left( { - l - i\left( {C_1  - C_2 } \right), - l - i\left( {C_1  + C_2 } \right);1 - 2iC_1 ;1} \right)
$. We should mention in finding the in-going solution, we consider the low frequency condition, $\omega  <  < 1/M$ in near region, $r <  < 1/\omega$, along with the assumption of small probe $eQ<<1$. 
Using the Gauss' theorem for Gamma functions, we can re-write the factor $A$ in equation (\ref{Rin-far}) as
\begin{align}
	A = \frac{{\Gamma \left( {1 - 2iC_1 } \right)\Gamma \left( {2l + 1} \right)}}{{\Gamma \left( {l + 1 - i\frac{{\left( {2M\omega  - m_\beta  Q\left( {1 - \beta } \right)} \right)\left( {r_ +   + r_ -  } \right) - 2m_\beta  a\beta }}{{r_ +   - r_ -  }}} \right)\Gamma \left( {l + 1 + i\left( {2M\omega  - m_\beta  Q\left( {1 - \beta } \right)} \right)} \right)}}.\label{AGauss}
\end{align}
Hence, we find the absorption cross section, given by 
\begin{align}
	P_{abs}  \sim \left| A \right|^{-2}  = \sinh \left( {2\pi C_1 } \right)\frac{{\left| {\Gamma \left( {l + 1 - iB_1 } \right)} \right|^2 \left| {\Gamma \left( {l + 1 - iB_2 } \right)} \right|^2 }}{{2\pi C_1 \left( {\Gamma \left( {2l + 1} \right)} \right)^2 }}\label{Pabs},
\end{align}
where
\begin{eqnarray}
B_1  &=& \frac{{\left( {2M\omega  - m' Q\beta} \right)\left( {r_ +   + r_ -  } \right) - 2m'  a\alpha }}{{r_ +   - r_ -  }},\\
B_2  &=& \left( {2M\omega  - m'  Q\beta} \right),
\end{eqnarray}
and $C_1$ and $C_2$ are given by (\ref{c1}) and (\ref{c1}), respectively. To find the possible agreement between macroscopic cross section $P_{abs}$ and the microscopic cross section of dual CFT, we need to identify some parameters of the theories. In this regard, we consider the first law of thermodynamics for the charged rotating black holes which can be written as
\begin{align}
	T_H \delta S_{BH} = \delta M - \Omega_H \delta J - V_H \delta ,Q\label{var-entro}
\end{align}
where $T_H$ and $\Omega$ are given by (\ref{TH}) and (\ref{Omeg}), and $V_H$ is the electrostatic potential. In the case of neutral rotating black holes, $\delta J$ can be identified as $m$ and $\delta M$ as $\omega$ \cite{KerrCFT1,KerrCFT2,KerrCFT3}. In addition to these identifications, for charged black holes we identify $\delta Q$ as $e$.

To find the conjugate charges, we calculate the
variation of entropy from gravitational point of view, $\delta S_{BH}$ as well as the variation of entropy from CFT, $\delta S_{CFT}$. These two variations should be equal, so we find
\begin{align}
	\frac{{\delta M - \Omega _H  \delta J - V_H \delta Q  }}{{T_H }} = \frac{{\delta E_L }}{{T_L }} + \frac{{\delta E_R }}{{T_R }}\label{bh-cft},
\end{align}
where $\Omega_H$ and $V_H$ are given by (\ref{Omeg}) and (\ref{electricpot}) respectively.
The absorption cross section (\ref{Pabs}) can be written as a thermal CFT absorption cross section if we identify  $ \delta E_L  = \tilde \omega _L $ and $\delta E_R  = \tilde \omega _R $ where 
\begin{align}
	\tilde \omega _L  = \frac{{\left( {2M\omega  - m'  Q\beta} \right)\left( {r_ +   + r_ -  } \right)}}{{2a\alpha }}\label{wL},
\end{align}
and
\begin{align}
	\tilde \omega _R  = \frac{{\left( {2M\omega  - m'  Q\beta} \right)\left( {r_ +   + r_ -  } \right)}}{{2a\alpha }} - m'\label{wR}. 
\end{align}
The variables $\tilde \omega_R$ and $\tilde \omega_L$ are introduced somehow to accommodate three sets of CFT parameters: the frequencies $\omega _{L,R}$, the charges $q_{L,R}$, and the chemical potentials $\mu_{L,R}$. The relations between these variables are written as
\begin{equation}
\tilde \omega _{L,R}  = \omega _{L,R}  - q_{L,R} \mu _{L,R}, \label{tildetonotilde}
\end{equation}
where
\begin{equation}
\omega _L  = \frac{{2M\omega \left( {r_ +   + r_ -  } \right)}}{{2a\alpha }} ~~~,~~~ \omega _R  = \omega _L  - m',\label{omegaLR}
\end{equation}
\begin{equation}
\mu _L  = \mu _R  = \frac{{Q\beta \left( {r_ +   + r_ -  } \right)}}{{2a\alpha }},\label{muLR}
\end{equation}
and $q_L = q_R = m'$. We also notice that for $\beta = 0$, i.e. the absence of left and right chemical potential, and $\alpha = 1$, the left and right frequencies (\ref{wL}) and (\ref{wR}) reduce to standard left and right frequencies for Kerr-Sen geometry with a chargeless test field \cite{KSChen}.
In fact, by equation (\ref{wR}), (\ref{wL}), and (\ref{intro17}), the macroscopic cross section (\ref{Pabs}) can be expressed as
\begin{align}
	P_{abs}  \sim T_L^{2h_L  - 1} T_R^{2h_R  - 1} \sinh \left( {\frac{{\omega _L }}{{2T_L }} + \frac{{\omega _R }}{{2T_R }}} \right)\left| {\Gamma \left( {h_L  + i\frac{{\omega _L }}{{2\pi T_L }}} \right)} \right|^2 \left| {\Gamma \left( {h_R  + i\frac{{\omega _R }}{{2\pi T_R }}} \right)} \right|^2 \label{final},
\end{align}
where we set $h_L  = h_R  = l + 1$. Equation (\ref{final}) is the well known finite temperature absorption cross section for a $2D$ CFT \cite{cms}.

\section{Generalized Hidden Conformal Symmetry with Deformation Parameter $\kappa$}
\label{sec:sec5kappa}
In sections (\ref{sec:sec3hidden}) and (\ref{sec:sec4:absorption}), we considered the propagation of a scalar field in the background of a generic non-extremal Kerr-Sen black hole and found evidences for a hidden conformal field theory in $\phi '$ picture. The metric function of Kerr-Sen black hole has two roots $r_+$ and $r_-$ where the scalar wave equation (\ref{eq5}) have poles in both locations. We may deform the wave equation (\ref{eq5}) near the inner horizon $r_-$ since for the non-extremal Kerr-Sen black hole we can consider $r$ to be far enough from $r_-$ such that the linear and quadratic terms in frequency which are coming from the expansion near the inner horizon can be dropped \cite{Lowe}. In this regard we consider the deformation of radial equation (\ref{eq7}) for the massless scalar filed by deformation parameter $\kappa$ as 
\begin{align}
	\left[ {\partial _r \left( {\Delta \partial _r } \right) + \frac{{\left( {2Mr_ +  \omega  - a_1 m'  } \right)^2 }}{{\left( {r - r_ +  } \right)\left( {r_ +   - r_ -  } \right)}} - \frac{{\left( {2M\kappa r_ +  \omega  - a_2 m'  } \right)^2 }}{{\left( {r - r_ -  } \right)\left( {r_ +   - r_ -  } \right)}}} \right]R\left( r \right) = l\left( {l + 1} \right)R\left( r \right),\label{KSdef}
\end{align}
where $a_1  = Qr_ +  \beta + a\alpha $ and $a_2  = Q\kappa r_ +  \beta + a\alpha $. The deformation parameter $\kappa$ and $r-r_-$ should satisfy $\kappa M^2a_2m'\omega << 2\sqrt{(M-b)^2-a^2}(r-r_-)$ as well as 
$\kappa^2 M^4 \omega^2 << 2\sqrt{(M-b)^2-a^2}(r-r_-)$ to drop the linear and quadratic terms in frequency from the expansion near the inner horizon pole while we still keep the near region geometry and low frequency scalar field as an electrically charged probe. 
We look now to a new set of vector fields that make $SL(2,\mathbb{R})$ algebra in such a way that the quadratic Casimir operator of the algebra represents the deformed  radial equation (\ref{KSdef}) of the scalar field. We consider the set of vector fields $L_{\pm}$ and $L_0$ given by 
\begin{align}
	L_ \pm   = e^{ \pm \rho t \pm \sigma \phi } \left( { \mp \sqrt \Delta  \partial _r  + \frac{{C_2  - \delta r}}{{\sqrt \Delta  }}\partial _\phi '   + \frac{{C_1  - \gamma r}}{{\sqrt \Delta  }}\partial _t } \right),\label{LpmLowe}
\end{align}
\begin{align}
	L_0  = \gamma \partial _t  + \delta \partial {_\phi '}, \label{L0Lowe}
\end{align}
which should satisy $
\left[ {L_ +  ,L_ -  } \right] = 2L_0 ,
\left[ {L_ \pm  ,L_0 } \right] =  \pm L_0 
$ as well as making the Casimir operator
\begin{align}
L_0 ^2  - \frac{1}{2}\left( {L_ +  L_ -   + L_ -  L_ +  } \right) = \partial _r \left( {\Delta \partial _r } \right) + \frac{{\left( {2Mr_ +  \omega  - a_1 m'  } \right)^2 }}{{\left( {r - r_ +  } \right)\left( {r_ +   - r_ -  } \right)}} - \frac{{\left( {2M\kappa r_ +  \omega  - a_2 m'  } \right)^2 }}{{\left( {r - r_ -  } \right)\left( {r_ +   - r_ -  } \right)}}.\label{casdef}
\end{align}

The coefficients of $\partial _r$ and $\partial _r^2$ in (\ref{casdef}) gives
two equations 
\begin{align}
	\rho C_1  + \sigma C_2  + M = 0\label{r1},
\end{align}
and
\begin{align}
	1 + \rho \gamma  + \sigma \delta  = 0\label{r2}.
\end{align}
Moreover, the coefficient of $\partial _{\phi '} ^2$ and $\partial _t^2$ yield
\begin{align}
	- \delta ^2 \left( {r - r_ +  } \right)\left( {r - r_ -  } \right) + C_2 ^2  - 2C_2 \delta r + \delta ^2 r^2  = a_1 ^2 \frac{{r - r_ -  }}{{r_ +   - r_ -  }} - a_2 ^2 \frac{{r - r_ +  }}{{r_ +   - r_ -  }}\label{m},
\end{align}
and
\begin{align}
	C_1 ^2  - \gamma^2 \left( {r - r_ +  } \right)\left( {r - r_ -  } \right) - 2C_1 \gamma r + \gamma ^2 r^2  = \frac{{4M^2 r_ +  ^2 }}{{\left( {r_ +   - r_ -  } \right)}}\left( {\left( {r - r_ -  } \right) - \kappa ^2 \left( {r - r_ +  } \right)} \right).\label{t}
\end{align}
The last possible term in (\ref{casdef}) that is proportional to the mixed derivative $\partial _\phi  \partial _t$ is
\begin{align}
	 - C_2 C_1  + \delta rC_1  - \delta r^2 \gamma  + \gamma \left( {r - r_ +  } \right)\left( {r - r_ -  } \right)\delta  + C_2 \gamma r =  - \frac{{2Mr_ +  }}{{\left( {r_ +   - r_ -  } \right)}}\left( {a_1 \left( {r - r_ -  } \right) - \kappa a_2 \left( {r - r_ +  } \right)} \right)\label{mt}.
\end{align}
The two different classes of solutions to equation (\ref{m}) (that we show by subscripts $a$ and $b$ are, 
\begin{eqnarray}
\delta_a  = \frac{{a_1  + a_2 }}{{r_ +   - r_ -  }}, \, && \, C_{2a}  = \frac{{a_1 r_ -   + a_2 r_ +  }}{{r_ +   - r_ -  }},\\
\delta_b  = \frac{{a_2  - a_1 }}{{r_ +   - r_ -  }}, \, && \, C_{2b}  = \frac{{a_2 r_ +   - a_1 r_ -  }}{{r_ +   - r_ -  }}.
\end{eqnarray}
These solutions substituted into equations (\ref{t}) and (\ref{mt}) give the corresponding $C_1$ and $\gamma$, that are given by
\begin{eqnarray}
\gamma_a  = \frac{{2Mr_ +  \left( {\kappa  + 1} \right)}}{{r_ +   - r_ -  }}, \, && \, C_{1a}  = \frac{{2Mr_ +  \left( {\kappa r_ +   + r_ -  } \right)}}{{r_ +   - r_ -  }}, \\
\gamma_b  = \frac{{2Mr_ +  \left( {\kappa  - 1} \right)}}{{r_ +   - r_ -  }}, \, && C_{1b}  = \frac{{2Mr_ +  \left( {\kappa r_ +   - r_ -  } \right)}}{{r_ +   - r_ -  }}.
\end{eqnarray}
So, the generators of $SL(2,\mathbb{R})$ for $a-$solutions are 
\[
L_ {\pm a}   = e^{ \pm \rho _1 t \mp \left( {\frac{{b\left( {1 + \kappa } \right)}}{{a\alpha \left( {1 - \kappa } \right)}} + 2\pi T_R } \right)\phi } \left[  \mp \sqrt \Delta  \partial _r  + \left( {Q\beta\frac{{r_ +  \left( {r_ -   + \kappa r_ +   - r\left( {1 + \kappa } \right)} \right)}}{{r_ +   - r_ -  }}} \right. \right.
\]
\begin{align}
\left. {\left. { - \frac{\alpha \Omega_H }{{2\pi T_H }}\left( {r - \left( {M - b} \right)} \right)} \right)\frac{{\partial _{\phi '}  }}{{\sqrt \Delta  }} + \left( {\frac{{r - r_ +  }}{{2\pi \Omega _H \alpha \left( {T_L  + T_R } \right)}} - \frac{{r - \left( {M - b} \right)}}{{2\pi T_H }}} \right)\frac{{\partial _t }}{{\sqrt \Delta  }}} \right],\label{LpmKSa}
\end{align}
and
\begin{align}
	L_{0a}  = \left( {\frac{1}{{2\pi T_H }} - \frac{1}{{2\pi \Omega _H \alpha \left( {T_L  + T_R } \right)}}} \right)\partial _t  + \left( {\frac{{Q\beta\left( {1 + \kappa } \right)}}{{8\pi M  T_H }} + \frac{{\Omega _H \alpha }}{{2\pi T_H }}} \right)\partial _{\phi '},\label{L0KSa} 
\end{align}
where
\[
\rho _1  \equiv \frac{b}{Mr_+{\left( {1 - \kappa } \right)}} + \frac{{Q\beta}}{{2Ma\alpha \left( {1 - \kappa } \right)}}\left( {M  \left( {1 + \kappa } \right) - \kappa r_ +  - r_-} \right).
\]
For the second class of solutions, we find
\[
L_ {\pm b}   = e^{ \pm \rho _2 t \mp \left( {\frac{b}{{a\alpha }} + 2\pi T_L } \right)\phi } \left[ { \mp \sqrt \Delta  \partial _r  + \left( {2Mr_ +  \Omega _H \alpha  + \frac{{Q\beta r_ +  \left( {\kappa r_ +   - r_ -   + r - \kappa r} \right)}}{{r_ +   - r_ -  }}} \right)\frac{{\partial _{\phi '}  }}{{\sqrt \Delta  }}} \right.
\]
\begin{align}
	\left. { + \left( {2Mr_ +   + \frac{{\left( {r - r_ +  } \right)}}{{2\pi \alpha \Omega _H \left( {T_L  + T_R } \right)}}} \right)\frac{{\partial _t }}{{\sqrt \Delta  }}} \right],\label{LpmKSb}
\end{align}
and
\begin{align}
	L_{0b}  = \left( {\frac{{ - 1}}{{2\pi \alpha \Omega _H \left( {T_L  + T_R } \right)}}} \right)\partial _t  + \left( {\frac{{Q\beta\left( { \kappa - 1 } \right)}}{{8\pi T_H M}}} \right)\partial _{\phi '}, \label{L0KSb} 
\end{align}
where
\[
\rho _2  \equiv \frac{{Q\beta\left( {r_ +  \left( {r_ -   - \kappa r_ +  } \right) + Mr_ +  \left( {\kappa  - 1} \right)} \right)}}{{2Mr_ +  a\alpha \left( {\kappa  - 1} \right)}} + 2\pi \alpha \Omega _H \left( {T_R  + T_L } \right).
\]
As we notice, the generators (\ref{LpmKSa}), (\ref{L0KSa}), (\ref{LpmKSb}) and (\ref{L0KSb}) of $SL(2,\mathbb{R})\times SL(2,\mathbb{R})$ reduce exactly to the generators of generalized hidden conformal symmetry of Kerr black hole  \cite{Lowe},
in the limit where  $\alpha  = 1$ and $b=0$. The left and right temperatures are given by $
T_L  = T_R \frac{{1 + \kappa }}{{1 - \kappa }}
$
and $T_R  = \frac{{r_ +   - r_ -  }}{{4\pi \alpha a}}$ respectively. This means the right temperature of generalized hidden CFT doesn't get any contribution from the deformation parameter $\kappa$ and so is the same as the right temperature of hidden CFT while the left temperature is affected by the deformation parameter $\kappa$. 
Demanding the agreement of microscopic entropy of CFT given by (\ref{intro30}) to the Bekenstein-Hawking entropy of Kerr-Sen black hole (\ref{entKS})
requires the central charges are given by
\begin{align}
	c_L  = c_R  = \frac{{6\left( {1 - \kappa } \right)a\alpha Mr_ +  }}{{\sqrt {\left( {M - b} \right)^2  - a^2 } }}. \label{c}
\end{align}
These central charges reduce to central charges of generalized hidden CFT of Kerr black hole where $\alpha=1$ and $b=0$.
As we mentioned earlier, the charged Gibbons-Maeda-Garfinkle-Horowitz-Strominger black hole is a special case of Kerr-Sen black hole when the rotational parameter is zero. In this limit, one can show the solutions to equations (\ref{r1}) and (\ref{r2}) exist only for special values of parameter $\kappa$.
In 
the b-branch, the solutions are $\sigma=0$ along with we get 
\begin{equation}
\rho  = \frac{{M - 2b}}{{4M\left( {M - b} \right)}},\kappa  =  - \frac{M}{{M - 2b}} \label{rhokappaGHS}.
\end{equation}
Consequently, the generators of $SL(2,\mathbb{R})$ for $b$-solutions (\ref{LpmKSb}),(\ref{L0KSb}) reduce to
\begin{equation}
L_{ \pm b}  = e^{ \pm \left( {\frac{{M - 2b}}{{4M\left( {M - b} \right)}}} \right)} \left( { \mp \sqrt \Delta  \partial _r  - 2\frac{{Q\beta \left( {M - r} \right)\left( {M - b} \right)}}{{\left( {M - 2b} \right)\sqrt \Delta  }}\partial _\phi   - 4\frac{{M\left( {M - r} \right)\left( {M - b} \right)}}{{\left( {M - 2b} \right)\sqrt \Delta  }}\partial _t } \right)\label{LpmGH},
\end{equation}
\begin{equation}
L_{0b}  =  - 4\frac{{M\left( {M - b} \right)}}{{M - 2b}}\partial _t  - 2\frac{{\beta Q\left( {M - b} \right)}}{{M - 2b}}\partial _\phi  \label{L0GH}.
\end{equation}
So, these are the generators of $SL(2,\mathbb{R})$ for the  
Gibbons-Maeda-Garfinkle-Horowitz-Strominger black hole. The generators (\ref{LpmKSa}) and (\ref{L0KSa}) of $SL(2,\mathbb{R})$ for $a$-solutions with $\kappa  = M\left( {M - 2b} \right)^{ - 1} $ give the same copy of generators as in (\ref{LpmGH}) 
and (\ref{L0GH}) with renaming the generators by $L_{\pm}\rightarrow -L_\mp, L_0\rightarrow -L_0$. We also note that generators (\ref{LpmGH}) 
and (\ref{L0GH}) in the special case of $Q=0$ reduce to the generators of $SL(2,\mathbb{R})_{Sch}$  for Schwarzschild black hole \cite{Bertinii}.

\section{Conclusions}
\label{sec:concl}

In this article, we extend the concept of Kerr/CFT correspondence in general picture for the non-extremal four dimensional Kerr-Sen black holes. In this regard, we find the hidden conformal symmetry in general picture for the solution space of charged scalar field.
The crucial assumption in this paper is that the charged scalar field in the background of Kerr-Sen is expected to reveal the CFT duals for each black hole hairs which posses $U(1)$ symmetry, such as angular momentum and charge. One of the main results of this paper is that there is no well defined Q picture for Kerr-Sen black hole in contrast to the well defined Q picture for the 
Kerr-Newman black hole \cite{Chen-JQpic,chen-SL2Z}. 
Moreover we can conclude the ``microscopic no hair conjecture'' proposed in \cite{Chen-JQpic} doesn't apply to Kerr-Sen black hole. This may be understood from the fact that the Kerr-Sen geometries are not obtained from Einstein-Maxwell theory. 
Moreover, the equation of motion for the dilaton in Kerr-Sen geometry is  different from the equation of motion for the Klein-Gordon field and this renders the possibility of writing the equation in terms of Casimir invariants (\ref{intro28}) of $SL(2,\mathbb{R})_R$ and $SL(2,\mathbb{R})_R$. This observation is in agreement with the fact that the non-gravitational fields don't contribute to the central charge of conformal field theory \cite{GH2}. 

The other main result of this paper is finding a family of generalized hidden conformal symmetry for the Kerr-Sen black hole in general picture. The family of generators is parameterized by the deformation parameter $\kappa$ in the radial equation of charged scalar field. In the special limit of the deformation parameter $\kappa$, the generalized hidden conformal symmetry reduces to a single copy of symmetry for the charged Gibbons-Maeda-Garfinkle-Horowitz-Strominger black hole.

It is an open question to find (if any) possible relation between these generalized hidden conformal symmetry and the closed conformal Killing-Yano two-form of Kerr-Sen black hole \cite{one1}.
\vspace{1cm} 
\bigskip

{\Large Acknowledgments}

This work was supported by the Natural Sciences and Engineering Research Council of Canada. HMS is supported by the Dean's scholarship from the college of graduate studies and research, University of Saskatchewan.\newline

{\large Appendix}\newline
The vectors 
$\partial _ + $, $\partial _ - $, and $\partial _ y $ in terms of coordinates $t,r$ and $\phi '$, can be written as
\begin{eqnarray}
 \partial _ +   = e^{ - \left( {2\pi T_R \phi '  + 2n_R t} \right)} \left( {\Delta ^{1/2} \partial _r  + Z_{\phi  + } \partial _{\phi '}   - Z_{t + } \partial _t } \right),~~~~~
\label{dplus}\\
\partial _y  = e^{ - \left( {\pi \left( {T_L  + T_R } \right)\phi '  + \left( {n_R  + n_L } \right)t} \right)} \left( {Z_{ry} \partial _r  + Z_{\phi y} \partial _{\phi '}   - Z_{ty} \partial _t } \right),~~~
\label{dy}\\
\partial _ -   = e^{ - \left( {2\pi T_L \phi  '  + 2n_L t} \right)} \left( {\Delta ^{1/2} \partial _r  + Z_{\phi  - } \partial _{\phi '}   - Z_{t - } \partial _t } \right).~~~
\label{dminus}
\end{eqnarray}
where
\[
 Z_{\phi  + }  = \frac{{\left( {n_R \left( {r_ +   - r_ -  } \right) - n_L \left( {r_ +   + r_ -  } \right) + 2n_L r} \right)}}{{4\pi \Delta ^{1/2} \left( {n_L T_R  - n_R T_L } \right)}} \]\[
 Z_{t + }  = \frac{{\left( {T_R \left( {r_ +   - r_ -  } \right) - T_L \left( {r_ +   + r_ -  } \right) + 2T_L r} \right)}}{{4\Delta ^{1/2} \left( {n_L T_R  - n_R T_L } \right)}} \]\[ 
 Z_{ry}  =  - \frac{{2\Delta }}{{\sqrt {\left( {r - r_ -  } \right)\left( {r_ +   - r_ -  } \right)} }} \]\[
 Z_{\phi y}  = \sqrt {\frac{{r_ +   - r_ -  }}{{r - r_ -  }}} \frac{{\left( {n_L  - n_R } \right)}}{{2\pi \left( {n_L T_R  - n_R T_L } \right)}} \]\[
 Z_{ty}  = \sqrt {\frac{{r_ +   - r_ -  }}{{r - r_ -  }}} \frac{{\left( {T_L  - T_R } \right)}}{{2\left( {n_L T_R  - n_R T_L } \right)}} \]\[
 Z_{\phi  - }  = \frac{{\left( {n_R \left( {r_ +   + r_ -  } \right) - n_L \left( {r_ +   - r_ -  } \right) - 2n_L r} \right)}}{{4\pi \Delta ^{1/2} \left( {n_L T_R  - n_R T_L } \right)}}
\]
and
\[
 Z_{t - }  = \frac{{\left( {T_R \left( {r_ +   + r_ -  } \right) - T_L \left( {r_ +   - r_ -  } \right) - 2T_R r} \right)}}{{4\Delta ^{1/2} \left( {n_L T_R  - n_R T_L } \right)}} \\ 
\]

\end{document}